\documentclass[11pt,a4paper]{article}

\usepackage[utf8]{inputenc}
\usepackage{tikz}
\usepackage{graphicx}
\usepackage{multirow}
\usepackage{amsmath,amssymb,amsfonts}
\usepackage{amsthm}
\usepackage{mathrsfs}
\usepackage[title]{appendix}
\usepackage{xcolor}
\usepackage{textcomp}
\usepackage{manyfoot}
\usepackage{booktabs}
\usepackage{algorithm}
\usepackage{algorithmicx}
\usepackage{algpseudocode}
\usepackage{listings}
\usepackage{braket}
\usepackage{array}
\usepackage{cancel}
\usepackage{url}

\newcommand{\be}{\begin{equation}}
	\newcommand{\ee}{\end{equation}}

\title{Classical Coherence Distinguishes Organisms from Colonies}

\author{Yehuda Roth}

\date{} 

\begin{document}
	
	\maketitle
	
	\setcounter{section}{0}
	\renewcommand{\thesection}{S\arabic{section}}
	\renewcommand{\theequation}{\thesection.\arabic{equation}}
	\numberwithin{equation}{section}
	\renewcommand{\thefigure}{S\arabic{figure}}
	\renewcommand{\thetable}{S\arabic{table}}
	\renewcommand{\thesubsection}{\thesection.\arabic{subsection}}

\begin{abstract}
	What distinguishes a multicellular organism from a mere colony? In organisms, cells lose functional autonomy and are defined primarily within the organismal context, whereas in colonies the collective behaves as a statistical ensemble of independently functioning units.
	
	In physics, coherence denotes a regime in which a many-body system is described by a single collective state rather than by independent degrees of freedom. By analogy, organismal unity motivates the hypothesis that an underlying form of coherence may operate in biological systems. Although quantum coherence has been proposed in biological settings, decoherence in warm, noisy environments occurs on picosecond timescales and is therefore unlikely to underlie developmental and organism-level phenomena. Here we introduce a concept of \emph{classical coherence} tailored to biological conditions.
	
	As a guiding analogy, the centre-of-mass coordinate represents a many-body system through a single collective variable, and identical particles localized within the same region can minimize energy in coherent configurations. Motivated by this structure, we explore the possibility that multicellular organisms exhibit a form of biological coherence that constrains cellular states at the organismal level, and we suggest specific DNA sequence configurations as one possible physical substrate of such coherence.
	
	To provide an empirical test, we propose an experiment using \emph{Dictyostelium discoideum}, which transitions between unicellular amoebae and a multicellular slug. By exposing both states to the same intracellular pathogen and measuring the distribution of infected cells—operationally defined via stable genetic or regulatory perturbations—we obtain a quantitative criterion to distinguish independent from coherent cellular responses. A deviation from the infection statistics expected under independence would provide evidence bearing directly on the proposed coherence framework.
\end{abstract}
\maketitle
	
%%%%%%%%%%%%%%%%%%%%%%%%%%%%%%%%%%%%%%%%%%%
\section{INTRODUCTION}
%%%%%%%%%%%%%%%%%%%%%%%%%%%%%%%%%%%%%%%%%%%
What distinguishes a living organism from a colony of cells?
A human body contains \(\sim 3 \times 10^{13}\) cells\cite{Sender2016}, yet infections
typically trigger an integrated immune response at the organismal scale, involving
coordinated activation of innate and adaptive immune networks across distant tissues%
~\cite{RoittImmunology,JanewayImmunobiology}. By contrast, a bacterial biofilm of comparable cell count responds as a stochastic aggregate\cite{Flemming2016,Nadell2016}. The cells may be structurally similar, the pairwise biochemical interactions nearly identical, yet one system behaves as \textit{one} and the other as \textit{many}. How does organismal unity emerge, and what distinguishes it from mere collectivity in physically precise terms?

We propose that this distinction reflects a fundamental physical property: \textit{coherence}. Following physics terminology, we define coherence as the property of many constituents being fully described by a single collective state\cite{Quantiki_separable,Hansenne2024operational}. In this language, a \textit{colony} is a multicellular assembly whose components remain dynamically independent—a separable product state. An \textit{organism}, by contrast, is one in which this independence is lost and the system resides in a single, non-separable collective mode. We therefore conjecture that as long as cells remain coherent, the organism behaves as a unified living system, whereas when coherence is lost, the system reduces to a mere collection of independent parts.

Standard explanations for organismal coordination invoke biochemical relay networks—cytokines, morphogens, and gap-junction coupling that propagate signals cell by cell. These mechanisms are well documented and explain many collective phenomena~\cite{Goldschmidt2014,Haken1983}. At the same time, several observations suggest coordination that is not easily reduced to local relay alone. Single-cell transcriptomics of antipathogen responses reveal cells separated by millimeters activating within time windows shorter than diffusive relay permits~\cite{Sun2020,Cheemarla2021}. Whole-organism metabolic rate can change abruptly at morphogenetic transitions~\cite{Gregor2010}. Multicellular organisms also consistently achieve more robust immune outcomes than isolated cell populations, even when individual defense molecules are comparable in efficacy. These phenomena naturally motivate a description in terms of global constraints on cellular states, rather than purely local signaling.

Quantum coherence has been proposed as a mechanism for fast, long-range coordination~\cite{Frohlich1968,Marais2018,Davies2021}, but decoherence in warm, wet environments is expected to destroy macroscopic entanglement on sub-picosecond timescales, far too brief for developmental or organismal processes~\cite{Marais2018}. We therefore seek a notion of coherence that is compatible with classical, noisy biological conditions.

Here we develop a framework in which coherence emerges from purely \textit{classical} dynamics. It has long been emphasized that correlated behavior among many particles is not exclusive to quantum mechanics: even in classical mechanics, a collection of particles can behave as a single effective unit when described in terms of its centre-of-mass coordinate. In such a description, the ensemble is represented by a coherent global degree of freedom \(Q\) that encodes the state of the whole, while the remaining internal coordinates capture incoherent fluctuations around this collective motion\cite{roth2019superposition}. We exploit this fact to argue that an analogous, classically styled coherence may exist in biological systems.

Although we do not specify the microscopic mechanism that generates such coherence, the close parallel between the definition of a multicellular organism and the physical notion of coherence—parts that have no independent meaning outside a single collective state~\cite{Quantiki_separable,Hansenne2024operational}—motivates us to posit a biologically unique form of classical coherence between cells. In the present work, we consider coherence between the DNA codes of different cells. Operationally, we say that DNA codes are coherent when identical sequences distributed across cells exist as a shared, non-separable configuration, rather than as independent, distinguishable copies. This means that an organism in a coherent state contains \(N\) copies of a given DNA sequence, but these copies collectively define the organism as a whole; individual assignments to specific cells become well defined only upon measurement. We stress that this is a classical, ensemble-level coherence, not a claim about quantum superposition of DNA molecules.

To test our coherence hypothesis, we focus on \textit{Dictyostelium discoideum}, a social amoeba whose cells can exist either as independent unicellular amoebae or as coordinated multicellular slugs~\cite{Kessin2001,Schaap2005,Schaap2016,dicty_lifecycle}. This natural ability to switch between colony-like and organism-like states under controlled conditions allows us to measure infection statistics in both regimes. Within our framework, a statistically significant overdispersion in the multicellular slug relative to the unicellular phase would constitute evidence for organism-level coherence, whereas the onset of individually countable infected cells would signal the collapse of coherence and, in this precise sense, the death of the organism.

The main aim of this paper is to identify a classical form of coherence in biological systems. This requires developing new mathematical frameworks. To ease the reader into the argument, we briefly outline the structure of the paper so that readers from different backgrounds can focus on the sections most relevant to them. In Sec.~\ref{sec:classical_coherence_fock} we develop a purely classical framework for coherence, using centre-of-mass dynamics and a Fock-like representation to formalise the distinction between coherent and separable many-body states. This section is mainly methodological and may be skipped by readers interested primarily in biological implications.

In Sec.~\ref{sec:infection_fock} we specialise the Fock-space description to cellular infection states (healthy versus infected cells) and define what we mean by a coherent organismal infection state versus a separable colony state. Sec.~\ref{supp:S2} introduces the restoration operator as a compact way to describe organism-wide repair and removal of infected cells within this coherent framework.

In Sec.~\ref{test} we translate these concepts into a concrete experimental proposal using \textit{Dictyostelium discoideum}, defining a simple statistical criterion based on the distribution of infected-cell counts to distinguish coherent organismal responses from independent colony behaviour. The Discussion section then interprets these results in biological terms and highlights how coherence provides a quantitative, testable notion of organismal unity.
%%%%%%%%%%%%%%%%%%%%%%%%%%%%%%%%%%%%%%%%%%%%%%%%%%%%%%%%%%%%%%%%%%%%%%
\section{Classical coherence}
\label{sec:classical_coherence_fock}
%%%%%%%%%%%%%%%%%%%%%%%%%%%%%%%%%%%%%%%%%%%%%%%%%%%%%%%%%%%%%%%%%%%%%%

The distinction between a multicellular organism and a colony --- parts that have no functional meaning outside a unified whole versus parts that remain autonomous --- motivates a search for a physical notion of coherence in purely classical systems. Our goal in this section is not to claim that organisms are literally described by a mechanical center of mass, but to show that classical mechanics already contains clear examples where many constituents are described by a single effective degree of freedom. This makes it logically consistent to postulate a non-quantum form of coherence in biology.

\subsection{Hamiltonian formulation and canonical structure}
\label{dis1}

We consider an ensemble of $N$ interacting units described by canonical variables $(q_i,p_i)$ with Hamiltonian
\begin{equation}
	H = \sum_{i=1}^{N} \frac{p_i^2}{2m_i} + V(q_1,\dots,q_N) .
\end{equation}
To expose the coherent structure, we perform a canonical transformation to collective and internal variables~\cite{Goldstein2002}. The global (coherent) coordinate is the centre of mass
\begin{equation}
	Q = \frac{1}{M} \sum_{i=1}^{N} m_i q_i , \qquad M = \sum_{i=1}^{N} m_i ,
\end{equation}
with conjugate momentum
\begin{equation}
	P = \sum_{i=1}^{N} p_i .
\end{equation}
The internal (incoherent) degrees of freedom are given by relative coordinates
\begin{equation}
	u_i = q_i - Q ,
\end{equation}
with conjugate momenta
\begin{equation}
	\pi_i = p_i - \frac{m_i}{M} P .
\end{equation}
These variables preserve the canonical structure, subject to the constraints
\begin{equation}
	\sum_{i=1}^{N} m_i u_i = 0 , \qquad \sum_{i=1}^{N} \pi_i = 0 ,
\end{equation}
ensuring that the internal sector contains $N-1$ independent degrees of freedom.

For potentials that can be decomposed as
\begin{equation}
	V(q_1,\dots,q_N) = V_{\text{ext}}(Q) + V_{\text{int}}(u_1,\dots,u_N) ,
\end{equation}
the Hamiltonian separates into coherent and incoherent sectors:
\begin{equation}
	H =
	\underbrace{\frac{P^2}{2M} + V_{\text{ext}}(Q)}_{H_{\text{coherent}}}
	+
	\underbrace{\sum_{i=1}^{N} \frac{\pi_i^2}{2m_i} + V_{\text{int}}(u_1,\dots,u_N)}_{H_{\text{incoherent}}} .
\end{equation}
Here $H_{\text{coherent}}$ describes the ensemble as an effective single classical degree of freedom, encoded by a low-dimensional collective coordinate $(Q,P)$, while $H_{\text{incoherent}}$ captures mass-weighted fluctuations around this global mode~\cite{roth2019superposition}. This use of the term ``coherent'' is purely classical and refers to dynamical alignment of many trajectories in phase space into a collective mode, not to quantum superposition or phase coherence.

%%%%%%%%%%%%%%%%%%%%%%%%%%%%%%%%%%%%%%%%%%%%%%%%%%%%%%%%%%%%%%%%%%%%%%%%%%%%%%%%%%%%%%%%%
\subsection{Identical particles at a common location as a prelude to biological systems}
%%%%%%%%%%%%%%%%%%%%%%%%%%%%%%%%%%%%%%%%%%%%%%%%%%%%%%%%%%%%%%%%%%%%%%%%%%%%%%%%%%%%%%%%%

We consider a system of \(N\) identical particles of mass \(m\), each subject to an identical harmonic potential with minimum at a common position \(a\). In the single-particle description the Hamiltonian reads
\begin{equation}
	H = \sum_{i=1}^{N} \left[ \frac{p_i^2}{2m} + \frac{1}{2}k\,(q_i - a)^2 \right] .
\end{equation}
Introducing the centre-of-mass coordinate and momentum
\begin{equation}
	Q = \frac{1}{N}\sum_{i=1}^{N} q_i , \qquad P = \sum_{i=1}^{N} p_i ,
\end{equation}
and relative coordinates and momenta
\begin{equation}
	u_i = q_i - Q , \qquad \pi_i = p_i - \frac{m}{M}P , \qquad M = Nm ,
\end{equation}
with the constraints \(\sum_{i} u_i = 0\) and \(\sum_{i} \pi_i = 0\), the Hamiltonian separates into a coherent centre-of-mass sector and an incoherent internal sector:
\begin{equation}
	H = H_{\text{coherent}}(Q,P) + H_{\text{incoherent}}(\{u_i,\pi_i\}) ,
\end{equation}
where
\begin{equation}
	H_{\text{coherent}}(Q,P)
	= \frac{P^2}{2Nm} + \frac{1}{2}\,kN\,(Q - a)^2 ,
\end{equation}
and
\begin{equation}
	H_{\text{incoherent}}(\{u_i,\pi_i\})
	= \sum_{i=1}^{N} \left[ \frac{\pi_i^2}{2m} + \frac{1}{2}k\,u_i^2 \right] .
\end{equation}

Suppose now that \(N\) identical particles are localized at a common position \(q=a\). In this case, the collective coordinate satisfies \(Q = a\), and the internal displacements vanish, \(u_i = 0\). At first glance this configuration appears indistinguishable from a single particle at \(q=a\), yet the system still contains many coherent particles governed by the full many-body Hamiltonian. In the perfectly coherent configuration
\begin{equation}
	q_i = a , \quad p_i = 0 \quad \forall i ,
\end{equation}
we have
\begin{equation}
	Q = a , \quad P = 0 , \qquad u_i = 0 , \quad \pi_i = 0 \quad \forall i ,
\end{equation}
so that the internal Hamiltonian vanishes,
\begin{equation}
	H_{\text{incoherent}} = 0 ,
\end{equation}
and the total energy reduces to that of a single collective degree of freedom,
\begin{equation}
	H_{\text{coh}}(Q,P) = \frac{P^2}{2Nm} + \frac{1}{2}kN\,(Q - a)^2 .
\end{equation}
Thus, when the collective coordinate sits at the minimum of the effective potential,
\begin{equation}
	Q = a , \quad P = 0 ,
\end{equation}
the configuration in which all units occupy the same coordinate \(q=a\) is simultaneously the minimum–energy configuration of the classical system. Crucially, this configuration still represents a genuinely many–body state: the system contains \(N\) identical particles whose dynamics are encoded in a single coherent mode \((Q,P)\), even though all internal degrees of freedom are frozen.

This scenario is rarely discussed in the \textit{classical literature}, because having \(N\) particles occupying exactly the same point in space seems unphysical in most classical contexts. However, in the biological analogy developed here, the role of the spatial coordinate \(q\) is played by a genetic coordinate: many cells distributed throughout the organism can share the same DNA sequence. In this sense, multiple “identical particles at the same location” correspond to multiple cells carrying an identical genetic sequence, which is entirely natural in a multicellular organism. The organism is therefore not a single cell, but a coherent many–cell system whose genetic state is fully captured by a single effective collective coordinate in genetic state space.
%%%%%%%%%%%%%%%%%%%%%%%%%%%%%%%%%%%%%%%%%%%%%%%%%%%%%%%%
\subsection{Classical stability of coherence}
%%%%%%%%%%%%%%%%%%%%%%%%%%%%%%%%%%%%%%%%%%%%%%%%%%%%%%%%%%%%

Quantum coherence has been proposed as a mechanism for fast, long-range biological
coordination\cite{Frohlich1968,Marais2018,Davies2021}. However, at physiological
temperatures, quantum superpositions involving many degrees of freedom are expected to
decohere on ultrafast timescales in warm, wet, and strongly dissipative environments,
making long-lived macroscopic quantum coherence at the organismal scale highly
unlikely\cite{Marais2018,Davies2021}. From a classical perspective, there is nevertheless
no need for long-lived quantum coherence in order to obtain a stable collective mode in a
noisy, dissipative environment. Driven–damped many-body systems routinely support robust
global degrees of freedom—such as centre-of-mass coordinates or synchronized oscillator
phases—that remain well defined despite continuous coupling to a thermal bath and internal
fluctuations\cite{Goldstein2002,Landau1980,Rayleigh1903,Langevin1908,roth2019superposition}.

The perfectly coherent configuration analysed in the previous subsection provides a simple
example: when the collective coordinate \(Q\) sits at a stable minimum of the effective
potential \(U_{\text{eff}}(Q)\) and all internal modes are frozen (\(u_i = \pi_i = 0\)),
the many-body system reduces to a single stable collective degree of freedom. More
generally, classical driven–damped systems can exhibit long-lived, low-dimensional
attractors in phase space that serve as classical analogues of coherence: they encode
global, system-level order parameters that remain well defined over time without requiring
quantum superposition.
%%%%%%%%%%%%%%%%%%%%%%%%%%%%%%%%%%%%%%%%%%%%%%%%%%%%%%%%%%%%%%%%%%%%%%
\subsection{Biological realization of classical coherence}
%%%%%%%%%%%%%%%%%%%%%%%%%%%%%%%%%%%%%%%%%%%%%%%%%%%%%%%%%%%%%%%%%%%%%%

The previous sections established a purely classical framework for coherence. In
particular, we analyzed the special case of \(N\) identical particles localized at the
same position \(q\), and showed that the collective coordinate collapses to this
common value, \(Q = q\), even though the underlying state remains genuinely many-body.

In our biological model, we replace the spatial coordinate \(q\) with a genetic
coordinate \(\iota\). Identical values of \(\iota\) correspond to identical DNA
codes, so that many cells sharing the same genetic sequence are the direct analogue
of many particles sharing the same coordinate. Under this mapping, the “common
coordinate” becomes a common DNA code built from the identical contributions of
each cell.

To formalize stability in this genetic state space, we introduce a classical
effective potential \(U_{\text{eff}}(\iota)\) on genetic coordinates, in analogy
with the use of scalar “landscapes” in evolutionary and systems
biology~\cite{Gavrilets2004,deVisser2014}. In fitness landscape models, local
optima of a genotype-to-fitness map correspond to genetically stable
configurations under selection. Here we use \(U_{\text{eff}}(\iota)\) in an
entirely classical sense: a genetic code \(\iota_0\) that minimizes
\(U_{\text{eff}}\) represents a dynamically stable attractor of the underlying
gene-regulatory dynamics. When many cells share this code \(\iota_0\), the
organism realizes a coherent global configuration that is robust to small
perturbations.

%%%%%%%%%%%%%%%%%%%%%%%%%%%%%%%%%%%%%%%%%%%%%%%%%%%%%%%%%%%%%%%%%%%%%%%%%%%
\section{Fock-space representation of cellular states under infection\label{sec:infection_fock}}
%%%%%%%%%%%%%%%%%%%%%%%%%%%%%%%%%%%%%%%%%%%%%%%%%%%%%%%%%%%%%%%%

To formalize organismal coherence, we represent the genetic state of each cell in a Fock-space framework. This allows us to distinguish between separable (colony) and non-separable (organism) configurations using a single state vector, analogous to quantum many-body systems but entirely classical.

Suppose that the host genetic code is denoted by $\checkmark$ and an intruder (pathogen) code by $\bcancel{\checkmark}$. 

We represent the cellular states using creation and annihilation operators $c^\dagger_{i,\sigma}$ and $c_{i,\sigma}$ (see Supplementary Information, Section~S2), satisfying the anticommutation relations
\begin{equation}
	\{c_{i,\sigma}, c^\dagger_{j,\tau}\} = \delta_{ij}\delta_{\sigma\tau}, \quad
	\{c_{i,\sigma}, c_{j,\tau}\} = 0,
\end{equation}
where $\sigma,\tau \in \{\checkmark, \bcancel{\checkmark}\}$. These relations ensure that each mode $(i,\sigma)$ can be either empty or occupied, $n_{i,\sigma} \in \{0,1\}$.

Retroviruses and other integrating viruses can insert their genetic material into the host genome and thereby durably modify the host DNA code\cite{Marais2018futureQB}. Likewise, many intracellular bacteria deliver and maintain their own genetic programs inside host cells, and their genomes are extensively shaped by horizontal gene transfer and mobile genetic elements\cite{LegionellaHGT,LegionellaIslands}. The mathematical formalism accommodates these possibilities as follows.

Each cell \(i\) has two Fock modes, \(\checkmark\) (host) and \(\bcancel{\checkmark}\) (intruder), with occupation numbers \(n_{i,\checkmark}, n_{i,\bcancel{\checkmark}} \in \{0,1\}\). This yields four basis states:
\begin{enumerate}
	\item Vacuum state: Empty reference state in which no modes are occupied\\
	\(
	|0\rangle_i = |0\rangle_{i,\checkmark}\, |0\rangle_{i,\bcancel{\checkmark}}.
	\)
	\item Healthy cell (host code only):\label{healthy}
	~~~~\(
	|\nu\rangle_i^{H} = |1\rangle_{i,\checkmark}\, |0\rangle_{i,\bcancel{\checkmark}}.
	\)
	\item Bacterially infected cell (obligate/facultative intracellular pathogen; both codes present in the same cell):\label{infected}
	~~~~\(
	|\bcancel{\nu}\rangle_i^{B} = |1\rangle_{i,\checkmark}\, |1\rangle_{i,\bcancel{\checkmark}}.
	\)
	\item Virally infected cell (virus code only):\label{vinfected}
	~~~~\(
	|\bcancel{\nu}\rangle_i^{V} = |0\rangle_{i,\checkmark}\, |1\rangle_{i,\bcancel{\checkmark}}.
	\) 
\end{enumerate}
\noindent
\textit{Thus, every cell is assigned to one of the biologically familiar categories (healthy, bacterially infected, virally infected), and the Fock-space notation simply provides a compact way to encode these possibilities for all cells at once.}
%%%%%%%%%%%%%%%%%%%%%%%%%%%%%%%%%%%%%%%%%%%%%%%%
\subsection{Colony versus organism: separable and non-separable states}
%%%%%%%%%%%%%%%%%%%%%%%%%%%%%%%%%%%%%%%%%%%%%%%%

\textbf{Colony (separable):} Each cell is in a definite state. The total state is a product:
\begin{equation}
	|N\rangle_{\text{col}} 
	= \prod_{i=1}^{N_\checkmark} |\nu\rangle_i 
	\prod_{i=N_\checkmark+1}^{N} |\bcancel{\nu}\rangle_i,
\end{equation}
where $N_{\checkmark}$ cells are healthy and $N_{\bcancel{\checkmark}} = N - N_\checkmark$ are infected. The number of infected cells is well-defined at each instant.

\textbf{Organism (non-separable):} We conjecture that organismal coherence is maintained as long as the organism is alive. Under this assumption, infection perturbs but does not immediately collapse the superposition:
\begin{equation}
	|N\rangle_{\text{org}} 
	= \sum_{\{\sigma_i\}} A_{\{\sigma_i\}} 
	\prod_{i=1}^{N} |\sigma_i\rangle,
	\label{supervirus}
\end{equation}
where $\sigma_i \in \{\nu_i, \bcancel{\nu}_i\}$ and the expansion coefficients $A_{\{\sigma_i\}}$ satisfy
\[
\sum_{\{\sigma_i\}} |A_{\{\sigma_i\}}|^2 = 1.
\]

We recall that $|\nu\rangle_i$ denotes cell $i$ carrying a single host code, with no intruder code, while $|\bcancel{\nu}\rangle_i$ denotes cell $i$ in which both codes are present in the same cell. 

The superposition of these states in Eq.~\eqref{supervirus} means that, before measurement, there is no single, definite number of healthy or infected cells: the organism is described by one coherent state that simultaneously encodes many possible cell-count configurations with different weights. Any measuring device, however, must return a single outcome, not a superposition of possibilities. To avoid a logical contradiction between a non-separable organismal state and a measurement apparatus that can only report one value, we are therefore forced, by consistency alone, to introduce a notion of collapse. In close analogy to quantum theory, we define that upon measurement the coherent organismal state is reduced to one definite state with one definite healthy and infected cell count.

Our proposed experiment is intended to serve as a touchstone test of this coherence and collapse hypothesis, as explained in Sec.~\ref{test}.

If the number of infected cells is not defined, the fight against the invader cannot be described as a cell-by-cell process but as a lateral, organism-wide response. In our framework, this means that the immune response acts coherently across the whole system, updating the shared state $\ket{N}_{\text{org}}$ rather than targeting individually specified infected cells. To formalize this coherent update, we now introduce the restoration operator.
%%%%%%%%%%%%%%%%%%%%%%%%%%%%%%%%%%%%%%%%%%%%%%%%%%%%%%%%%%%%%%%%%%%%%%
\section{Restoration operator}
\label{supp:S2}
%%%%%%%%%%%%%%%%%%%%%%%%%%%%%%%%%%%%%%%%%%%%%%%%%%%%%%%%%%%%%%%%%%%%%%

Before we continue, we clarify our notation.
A healthy single cell is denoted by
\[
|\nu\rangle_i^H = |1\rangle_{i,\checkmark}|0\rangle_{i,\bcancel{\checkmark}}\label{healthy},
\]
while a sick single cell is denoted by
\[
|\bcancel{\nu}\rangle_i^B = |1\rangle_{i,\checkmark}\,|1\rangle_{i,\bcancel{\checkmark}}.
\]
Both of these are elements of the (classical) Fock space.

At the single-cell level, we define the states $|\checkmark\rangle$ and $|\bcancel{\checkmark}\rangle$ to represent, respectively, a cell with a correct (host) code and a cell carrying an additional intruder code in the underlying Hilbert space.

In biological terms, immune responses to infection are often described as the outcome of many local decisions, with each cell independently choosing whether to activate defense programs or undergo apoptosis~\cite{immune_system_review,immune_system_review2}. In our coherence-based framework, by contrast, the fight against infection is understood as a global process: the organism is described by a single coherent state, and the response to infection corresponds to an update of this collective state rather than to a sum of independent cellular actions. To represent the fact that many cells are restored simultaneously, we introduce a \textit{restoration operator} that drives the system back toward its healthy genetic--regulatory configuration. Operationally, this operator decreases the occupation of ``infected'' cellular states and increases the occupation of ``restored'' states, but it does so at the level of the shared state vector. In this way, the organism's coordinated defense against infection appears as a single transformation of a coherent state, rather than as a collection of unrelated microscopic events~\cite{Davies2021physicsbio,Marais2018futureQB}.

Biologically, host defense against infection does not begin with the destruction of infected cells but with an attempt to repair and restore them. Cells deploy a rich repertoire of repair mechanisms, including DNA damage repair pathways, protein quality-control systems, and organelle turnover, all aimed at returning the perturbed cell to a functional state within the organism~\cite{immune_system_review,immune_system_review2}. In our framework, these processes are captured by a \emph{repair channel} of the restoration operator, which maps infected cellular configurations back to healthy ones by transferring weight from ``infected'' to ``restored'' occupancy states while keeping the cell within the coherent ensemble.

Only when damage exceeds reparable thresholds, or when the infected cell poses a danger to the organism as a whole, does a \emph{destruction channel} become dominant. This channel removes severely perturbed cells from the coherent state, transferring their weight to an absorbing ``removed'' sector that no longer participates in $|\Psi\rangle$~\cite{Davies2021physicsbio,Marais2018futureQB}. Both channels thus contribute to preserving organismal coherence: repair acts to reinstate perturbed cells into the coherent state, whereas destruction sacrifices individual cells in order to protect the integrity of the coherent whole.

\paragraph{Repair in Hilbert space, destruction in Fock space.}
In our framework, repair and destruction act at different levels.
Repair is represented as a transformation within the single-cell Hilbert space:
\begin{equation}
	\mathbb{R}_r = \sum_i |\checkmark\rangle_i \langle \bcancel{\checkmark}|_i ,
\end{equation}
which maps an infected cell state to a healthy one without removing the cell from the ensemble; the cell remains part of the organismal Fock space, but its code is changed.

In contrast, destruction is defined at the Fock-space level, where an entire cell mode is removed from the coherent ensemble. Symbolically, we write
\begin{equation}
	\mathbb{R}_d = \sum_i c_{i,\bcancel{\checkmark}},
\end{equation}
where $c_{i,\bcancel{\checkmark}}$ is the annihilation operator associated with an infected cell mode. This operator reduces the number of participating cells by eliminating severely perturbed cells from the organismal state.

In practice, both channels act in parallel during host defense: $\mathbb{R}_r$ implements code-level repair of infected cells, while $\mathbb{R}_d$ removes cells that cannot be safely restored. Together they define the effective restoration dynamics of the coherent state.

%%%%%%%%%%%%%%%%%%%%%%%%%%%%%%%%%%%%%%%%%%%%%%%%%%	
\section{Testing the coherence hypothesis\label{test}}	
%%%%%%%%%%%%%%%%%%%%%%%%%%%%%%%%%%%%%%%%%%%%%%%%%%%

The superposition of cell configurations expressed in Eq.~\ref{supervirus} implies that the number of infected cells is not well-defined in the coherent phase. However, when an observer measures the number of infected cells, we conjecture that a collapse occurs, projecting the superposition onto a definite configuration with a definite value of $N_{\bcancel{\checkmark}}$. This leads us to propose the following experiment to probe our coherence hypothesis.

A natural experimental system is the social amoeba \textit{Dictyostelium discoideum}, whose cells can exist either as autonomous unicellular amoebae (colony phase) or as multicellular slugs (organism phase).\cite{Kessin2001,Schaap2016} In nutrient-rich conditions, \textit{Dictyostelium} behaves as a collection of independent cells; upon starvation, the same cells aggregate into a multicellular structure with a well-defined developmental program.

We consider intracellular pathogens that introduce an additional genetic code into each host cell, so that an infected cell carries both the host code and an intruder code. In our Fock-space representation, this corresponds to a double-occupancy state $|\bcancel{\checkmark}\rangle_i = |1\rangle_{i,\checkmark}\,|1\rangle_{i,\bcancel{\checkmark}}$, where the intruder code represents a stable genetic--regulatory perturbation of the cell's configuration.

As a concrete example, intracellular bacteria such as \textit{Legionella pneumophila} replicate within amoebae and introduce an additional genetic program into each host cell. \textit{Legionella} uses its Icm/Dot type IV secretion system to deliver hundreds of effector proteins into the host cytosol and to remodel host pathways, and its own genome is shaped by extensive horizontal gene transfer and mobile genomic islands that integrate into chromosomal DNA.\cite{LegionellaEffectors,LegionellaHGT,LegionellaIslands} In our language, a \textit{Dictyostelium} cell infected by \textit{Legionella} carries both the host code and an intruder code, and thus realizes the double-code state $|\bcancel{\checkmark}\rangle_i$. Infection could be read out, for example, by a fluorescent reporter or by single-cell assays that detect the presence of the intruder code.

\textbf{Proposed experiment:} 
We prepare two groups of \textit{Dictyostelium discoideum} with comparable total mass and cell density:
\begin{itemize}
	\item Group A (organism): multicellular slugs in the coherent phase.
	\item Group B (colony): unicellular amoebae in the autonomous phase.
\end{itemize}
Both groups are exposed to the same intracellular pathogen that introduces an intruder code into host cells at a controlled multiplicity of infection. After a fixed time interval, the number of cells carrying the intruder code $N_{\bcancel{\checkmark}}$ is measured in each sample using a consistent single-cell readout (for example, flow cytometry or imaging-based counting of reporter-positive cells). In practice, many replicate slugs and amoebal cultures would be infected under identical conditions, and $N_{\bcancel{\checkmark}}$ would be quantified separately in each replicate. Our coherence hypothesis is specifically about the variability across these replicates: it predicts a systematically broader distribution of $N_{\bcancel{\checkmark}}$ across organismal samples than across colony samples. In statistical terms, we expect overdispersed infection statistics in the multicellular slugs relative to the unicellular amoebae, reflecting the collapse of a non-separable state rather than a product of independent cell states.
\section{Discussion and outlook}
The manuscript has developed a framework in which coherence and collapse appear as global features of many-body systems that need not rely on quantum mechanics. In the present approach, coherent states are most naturally described in terms of a single collective mode, with individual constituents (cells, electrons, photons) treated as parts of this mode rather than as independently addressable units. When such a non-separable description is probed by a measurement device that must return a definite outcome, the act of measurement effectively selects one concrete configuration with well-defined individual values. In this view, collapse can be interpreted as a pragmatic resolution of the mismatch between a non-separable description and an apparatus constrained to yield a single result, and similar consistency requirements may operate in both quantum and classical coherent systems.

From a biological perspective, the starting point is the empirical contrast between multicellular organisms and colonies. In colonies, individual cells retain autonomous function and the aggregate behaves as a statistical ensemble of independently acting units. In multicellular organisms, by contrast, cells have no functional meaning except as parts of a unified whole, and infection often triggers coordinated responses that are difficult to explain purely in terms of local signalling and independent cellular decisions. We translated this intuition into a classical Fock-space description in which organismal cells occupy a non-separable state over genetic configurations, while colony cells occupy separable product states, and we showed how this leads to a concrete prediction about infection statistics that can be tested in \textit{Dictyostelium discoideum}.

Biologically, the main implication of the coherence hypothesis is that the number of infected (or healthy) cells in a coherent organism need not be a well-defined quantity prior to observation. Instead, the organism is described by a single state that encodes many possible infection configurations, and measurement selects one of them. Operationally, this leads us to expect broader variance and overdispersion in infected-cell counts across coherent multicellular slugs than across unicellular amoebae exposed to the same pathogen. In addition, the restoration operator provides a compact way to discuss immune repair and cell removal as updates of a global organismal state rather than as independent, cell-by-cell events, offering a complementary language for phenomena such as robustness of immune responses and organism-wide maintenance of function.

For readers with a physics background, our results suggest that several conceptual roles often associated with quantum coherence---loss of separability, global modes, measurement-induced selection of outcomes---can also be realised within a purely classical setting. The centre-of-mass analogy illustrates how a low-dimensional collective degree of freedom can encode the behaviour of many constituents, and the classical Fock-space construction extends this idea to discrete cellular configurations. In this sense, multicellular organisms provide a natural arena in which classical non-separability is not only conceptually meaningful but also accessible to empirical probing, with infection statistics serving as a many-body diagnostic.

In the multicellular context addressed here, we have modelled coherence as arising from DNA sequence configurations distributed across cells in a non-separable state, sustained by metabolism and lost at organismal death. This should be viewed as a working hypothesis rather than a definitive statement about the unique substrate of coherence in biology. The underlying idea is more general: any set of cellular variables capable of supporting a dynamically maintained collective mode could in principle realise classical coherence. If experiments in \textit{Dictyostelium} provide support for this picture, it would be natural to explore extensions of the framework to unicellular systems, to intracellular structures such as plasmids or mRNA copy numbers, and to other forms of biological organisation. More broadly, the present results indicate that coherence may contribute to physically grounded criteria for organismal unity, and they suggest a possible common language for future work at the interface of physics and biology.
\section*{Acknowledgments}
The author acknowledges the use of AI-assisted tools (Perplexity) for literature search, manuscript editing, and \LaTeX{} formatting. All scientific concepts, mathematical derivations, interpretations, and conclusions are the author's own.\\[0.5em]
The author would also like to thank Professor Yoram Gershman for helpful advice and discussions.
%\bibliographystyle{unsrt}  
%\bibliography{sources}

\end{document}